\documentclass[10pt,letterpaper]{article}
\usepackage{opex3}
\usepackage{color}
\usepackage{threeparttable}

\begin{document}

\title{A high-Q, ultrathin-walled microbubble resonator for aerostatic pressure sensing}

\author{Yong Yang,$^{1,2}$ Sunny Saurabh,$^1$ Jonathan M. Ward,$^1$ and S\'ile Nic Chormaic$^{1,*}$}

\address{$^1$Light-Matter Interactions Unit, Okinawa Institute of Science and Technology\\ Graduate University, Onna, Okinawa 904-0495, Japan\\
$^2$National Engineering Laboratory for Fiber Optics Sensing Technology, Wuhan University of Technology, Wuhan, 430070, China\\
}

\email{$^*$sile.nicchormaic@oist.jp} 



\begin{abstract}
Sensors based on whispering gallery resonators have minute footprints and can push achievable sensitivities and resolutions to their limits. Here, we use a microbubble resonator, with a wall thickness of 500 nm and an intrinsic Q-factor of $10^7$ in the telecommunications C-band, to investigate aerostatic pressure sensing via stress and strain of the material.   The microbubble is made using two counter-propagating CO$_2$ laser beams focused onto a microcapillary.  The measured sensitivity is 19 GHz/bar at 1.55 $\mu$m. We show that this can be further improved to 38 GHz/bar when tested at the 780 nm wavelength range. In this case, the  resolution for pressure sensing can reach 0.17 mbar with a Q-factor higher than $5\times10^7$.
\end{abstract}


\bibliography{SensePressure}
\bibliographystyle{osajnl}
\section{Introduction}
Whispering gallery mode (WGM) microresonators (WGRs) have been used as high sensitivity sensors for several years, with a wide variety of applications already reported,  such as temperature \cite{Dong2009} and bio/chemical \cite{Vollmer2008} sensing. With optical quality (Q) factors greater than $10^5$, WGM-based sensors  usually offer a high resolution and this has been already exploited for single nanoparticle detection \cite{Zhu2009, Li2014}.

Typically, WGRs are optically coupled using external  waveguides, such as tapered optical fibers. The resonator-waveguide sensing system is often placed into an aqueous environment, although, ideally, the coupling waveguide should be isolated from the fluid. For this purpose, hollow WGR structures based on a microcapillary \cite{Sumetsky2010,Li2010, Han2014APL, Watkins2011} have been developed. For a review on this topic the reader is referred to \cite{JWard2014}.  In contrast with solid WGRs, sensing can occur at the inner surface of the structure, while the WGMs are still excited by the external coupler via evanescent fields. For such structures, liquid can flow through the microcapillary and the fluid's properties, such as refractive index \cite{Li2010} or viscosity \cite{Han2014APL}, can be measured without disturbing the coupler. In general, there are two types of hollow WGR structures: (i) the microcapillary or bottle-like microcapillary resonator \cite{Li2010, Han2014APL} and (ii) the microbubble resonator \cite{Sumetsky2010, Watkins2011}. Currently, the highest Q-factors reported for the first type of hollow WGR are less than $10^5$, for a wall thickness of 2 $\mu$m \cite{Li2010}. The second type of device, i.e. the microbubble \cite{Sumetsky2010,Watkins2011}, is considered to have higher Q due to its curvature geometry which confines the WGMs more tightly in the polar direction compared with the first category of hollow resonators \cite{li2010analysis, Yang2014}. A few applications have been reported in microbubble WGRs, such as strain \cite{Sumetsky2010}, bio/chemical \cite{Sun2011} and temperature \cite{Ward2013} sensing, and lasing \cite{Lee2011}. For improving sensitivity in a bubble or capillary filled with liquid, the device must operate in the \textit{quasi-droplet regime} \cite{Yang2014, Lee2011}. This means that both a thin wall and a high Q-factor must be achieved for the microbubble.

In this paper, we report on the fabrication of ultrathin-walled (approaching 500 nm), silica microbubbles with Q-factors of about $10^7$ at the telecommunications C-band. By using the finite element method, we show that the obtained Q-factor approaches the theoretical upper bound set by the microbubble diameter and wall thickness.

\section{Measured Q-factor of the ultrathin-walled microbubble}
\begin{figure}
\centering\includegraphics[width=.7\columnwidth]{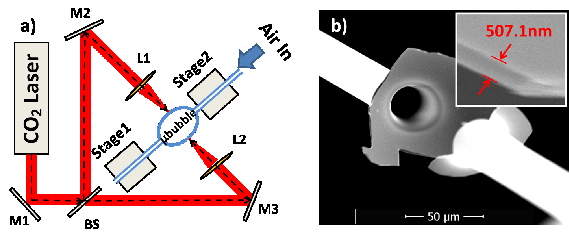}
\caption{(a) Schematic of the microbubble fabrication setup. M:mirror, L:CO$_2$ lens, BS:beamsplitter; (b) An SEM image of a microbubble snapped in the middle. This is Sample A in the main text.  The measured minimum wall thickness is 507 nm.}
\label{fab}
\end{figure}
\begin{figure}[h]
\centering\includegraphics[width=.7\columnwidth]{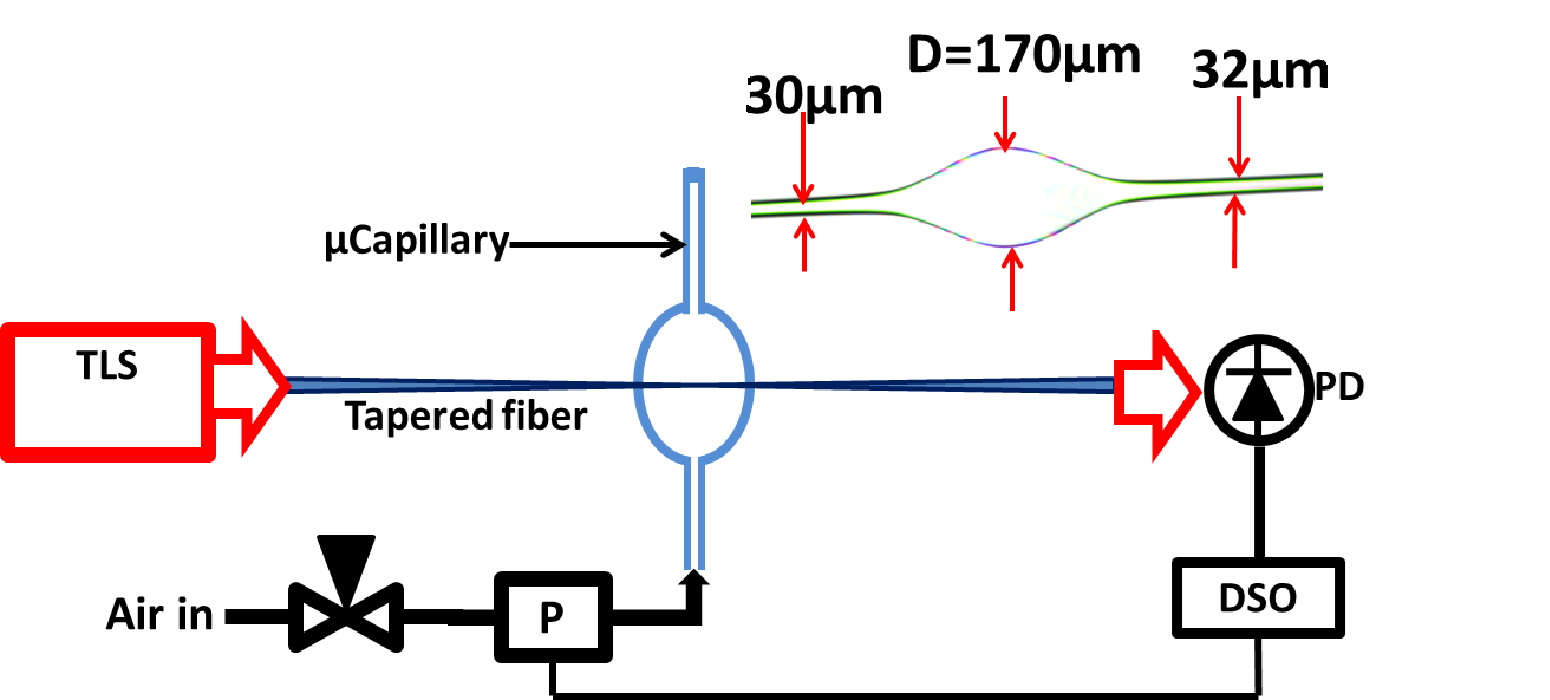}
\caption{Experimental setup for aerostatic pressure sensing. P:electronic pressure sensor, PD:photon detector, DSO:digital oscilloscope, TLS:tunable laser source.}
\label{setup}
\end{figure}
\begin{figure}[h]
\centering\includegraphics[width=.70\columnwidth]{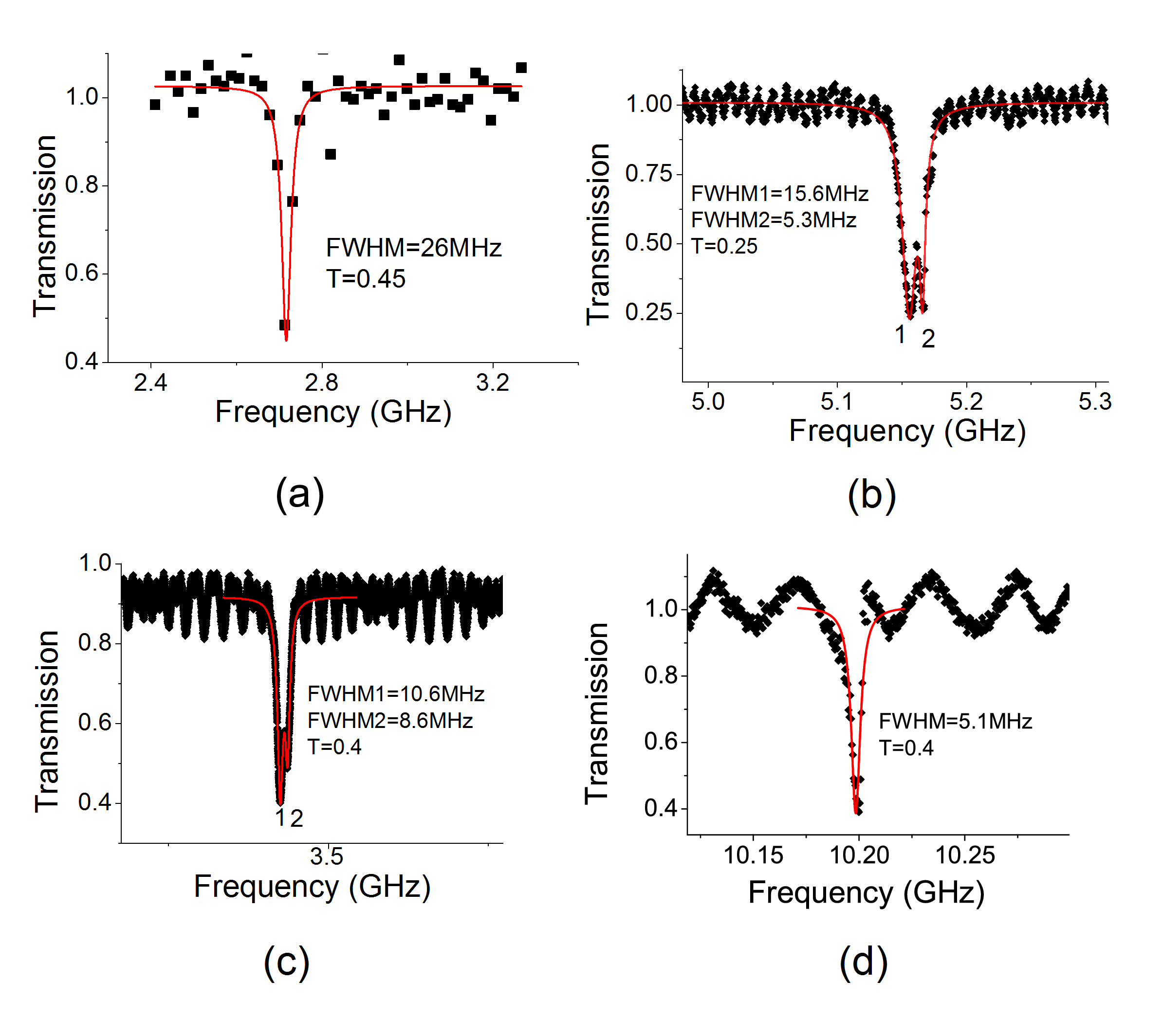}
\caption{ Q-factor measurements for two microbubble samples having similar geometrical parameters, A (upper) and B (lower). (a) Sample A at 1.55 $\mu$m; (b) Sample A at 780 nm; (c) Sample B at 1.55 $\mu$m; (d) Sample B at 780 nm. Note that modal coupling occurs for Sample A at 780 nm and for Sample B at 1.55 $\mu$m.}
\label{qmes}
\end{figure}

We attribute the high Q-factor that we measured to the fabrication process used, whereby two counter-propagating CO$_2$ laser beams are focused on the capillary, as depicted in Fig. \ref{fab}(a). This setup is inherently more stable than fabrication that relies on rotation of the capillary \cite{Sumetsky2010} or the heat source \cite{Berneschi2011}, and produces bubbles with a more uniform wall thickness. The crucial step in achieving an ultrathin microbubble is to use a silica capillary with a sufficiently small outer diameter and an inner-outer diameter ratio of 0.7 or greater. For example, in the setup described here, a capillary with an initial ID of 250 $\mu$m and OD of 350 $\mu$m was chosen and then tapered down to an OD of around 30 $\mu$m. To taper the capillary, each end was clamped onto a translation stage and the CO$_2$ beam was focused onto it. As the laser heated the capillary, one stage pulled the capillary out of the laser focus while the other fed the capillary into the focus at a much lower speed. This produced a thin microcapillary with a uniform waist at the required diameter. The microcapillary was next filled with compressed air and a microbubble was formed at the laser focus. The expansion stops at a certain wall thickness and, by increasing the CO$_2$ laser power, larger diameter (i.e. thinner walled) dual-input microbubbles are formed. By controlling the power and heating time, repeatable microbubbles of diameters around 170 $\mu$m and wall thicknesses between 507-570 nm were fabricated (Samples A and B). The wall thickness was determined from scanning electron microscopy (SEM) images.  As an example  see Fig. \ref{fab}(b) for Sample A. In the following we assume an average wall thickness of 550 nm. The Q-factor was measured by non-contact taper fiber coupling and laser scanning using either a 780 nm or 1550 nm tunable laser source, depending on the experiment.  The experimental setup is illustrated in Fig. \ref{setup}. Typical measurement results for two samples made using the same fabrication process, labeled Sample A and Sample B, are shown in Fig. \ref{qmes}. The intrinsic Q-factor can be deduced from the known transmission efficiency through the coupler, $T$, and the measured loaded Q.  Note that we have assumed ideal undercoupling. From the data in Fig. \ref{qmes}(a), the intrinsic Q is $9\times10^6$ for Sample A at 1.55 $\mu$m. At 780 nm, double peaks appear due to modal coupling, as shown in Fig. \ref{qmes}(b). The highest intrinsic Q-factor for peak 2 is $1\times10^8$.

\begin{figure}[h]
\centering\includegraphics[width=0.45\columnwidth]{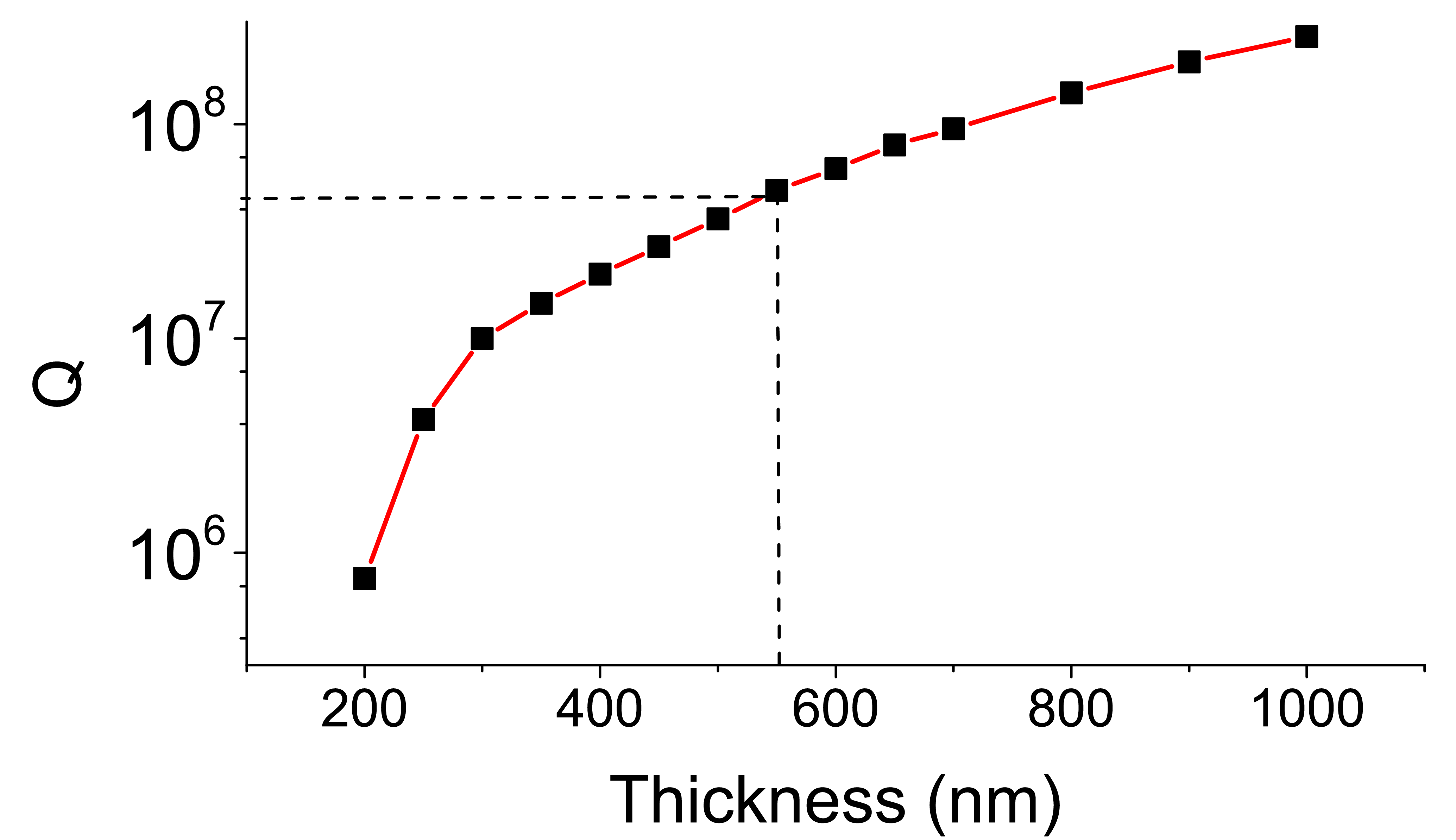}
\caption{Theoretical Q-factor of a 170 $\mu$m microbubble calculated for different wall thicknesses using a finite element method at 1.55 $\mu$m. The dashed line represents a wall thickness of 550 nm. The red line is a guide for the eye.}
\label{qcal}
\end{figure}

The intrinsic Q-factor of a WGR arises from the radiation and material loss. For solid WGRs, radiation loss is due to the evanescent field leaking at the medium boundary. Since for air-filled microbubbles both the inner and outer surfaces can be treated as leaky boundaries, the total radiation loss is larger than for solid WGRs. To accurately estimate the radiation loss, a modified finite element method \cite{Yang2014} can be used. The Q-factor values for a microbubble with a diameter of 170 $\mu$m at 1.55 $\mu m$ as a function of wall thickness is calculated and shown in Fig. \ref{qcal}. The material loss was also included, so that the maximum total Q-factor does not go beyond $10^9$. In practice, we would expect the Q-factor to be less than the theoretical predictions. In \cite{Yang2014}, simulations show that the sensitivity is exponentially (or near exponentially) related to the  wall thickness, $t$, and that the Q-factor also increases exponentially as a function of the outer radius, $R$. Thence, we can define two figures-of-merit (FOM) for comparison studies between different resonators, such that FOM$_1=Q\times 10^{\lambda/t}$ for the wall thickness and FOM$_2=Q\times 10^{\lambda/R}$ for the radius. The working wavelength, $\lambda$, is also taken into account.

Experimentally, a number of methods have been used for making thin-walled microbubbles \cite{Sumetsky2010, Li2010, Lee2011, Berneschi2011, Henze2011}. To summarize these experiments, in Table.\ref{tab1} we present the FOMs for the different methods for a direct comparison.
\begin{table}[h]
\centering
\begin{threeparttable}
\caption{\label{tab1}FOMs FOR thin-walled microbubbles for different fabrication methods.}
\begin{tabular}{|l|l|l|l|l|l|l|}
\hline
\textit{Reference} & \textit{Method} & $\lambda$ (nm) & $R$ ($\mu$m) & t ($\mu$m) & FOM$_{1}$ & FOM$_{2}$ \\ \hline
Berneschi \textit{et al.} \cite{Berneschi2011} & Electrical arc & 1550 & 170 & 4 & $1.4\times10^8$ & $6\times10^7$ \\ \hline
Sumetsky \textit{et al.} \cite{Sumetsky2010} &CO$_2$ with rotation& 1550 & 185 & 2 & $9\times10^6$ & $1.5\times10^6$  \\ \hline
Lee \textit{et al.} \cite{Lee2011}$^a$ &CO$_2$ with rotation & 780 & 102 & 0.56 & $2.5\times10^7$ & $1\times10^6$ \\ \hline
Li \textit{et al.} \cite{Li2010}$^a$ & Chemical etching & 980 & 35 & 1 & $1.2\times10^6$ & $1.3\times10^5$ \\ \hline
Henze \textit{et al.} \cite{Henze2011} & Two CO$_2$ beams & 780 & 192 & 2.9 & $1.8\times10^4$ & $1\times10^4$ \\ \hline
Sample A & Two CO$_2$ beams & 780 & 85 & 0.5 & $3.7\times10^9$ & $1\times10^8$ \\ \hline
Sample A & Two CO$_2$ beams & 1550 & 85 & 0.5 & $1\times10^{10}$ & $9.4\times10^6$ \\ \hline
\end{tabular}
\begin{tablenotes}
\item [] $^a$ Liquid-filled microbubble
\end{tablenotes}
\end{threeparttable}
\end{table}
With the fabrication method reported in \cite{Henze2011} and improved here, the FOMs increase 10-100 times for both the wavelengths, 1.55 $\mu$m and 780 nm, compared to devices fabricated using other methods. For the geometrical parameters of the microbubbles we fabricated, the theoretical, intrinsic Q-factor is $4.9\times10^7$ at 1.55 $\mu$m, shown in Fig. \ref{qcal}. The actual measured Q-factor approaches this theoretical limit within a factor of 5. A larger Q-factor implies that the sensing resolution for microbubbles could be improved. In the following section, we will demonstrate high resolution aerostatic pressure sensing.

\section{Resolution for aerostatic pressure sensing}
By applying aerostatic pressure to a WGR, the frequency of the WGM can shift by two means: (i) a stress-induced refractive index change and (ii) a strain-induced size expansion. Lavin \textit{et al.} \cite{martin2013high} investigated a doped, barium silicate solid microsphere for high pressure sensing. The achieved sensitivity was 0.0225 GHz/bar at a wavelength of 880 nm. Due to the low sensitivity from the index change, the minimal detectable pressure was $5\times10^3$ bar. In a liquid-formed droplet microsphere, which has a higher elasto-optical coefficient, the sensitivity can be significantly improved. However, the Q-factor is low ($10^4$) due to material absorption, thus limiting the resolution \cite{weigel2012}.

In a hollow microspherical structure, strain plays a dominant role \cite{ioppolo2007pressure, Yang2015}, and it can lead to an increase in sensitivity to aerostatic pressure. In polymethyl methacrylate (PMMA) hollow microspheres \cite{ioppolo2007pressure}, the sensitivity to the external pressure was improved to 3.44 GHz/bar, with a Q-factor up to $10^6$. In 2011, Henze \textit{et al.} demonstrated  internal pressure tuning in a silica microbubble \cite{Henze2011}. For a wall thickness of 2.9 $\mu$m and a diameter of 384 $\mu$m, the sensitivity was measured to be 22 GHz/bar at a wavelength of 780 nm. The Q-factor was about $10^4$ and this low value was attributed to  deformation during the fabrication process. Here, we show that both the sensitivity and Q-factor can be preserved even when the wall is ultrathin, thus enabling us to improve the resolution by a factor of almost 1000.

\begin{figure}
\centering\includegraphics[width=0.85\columnwidth]{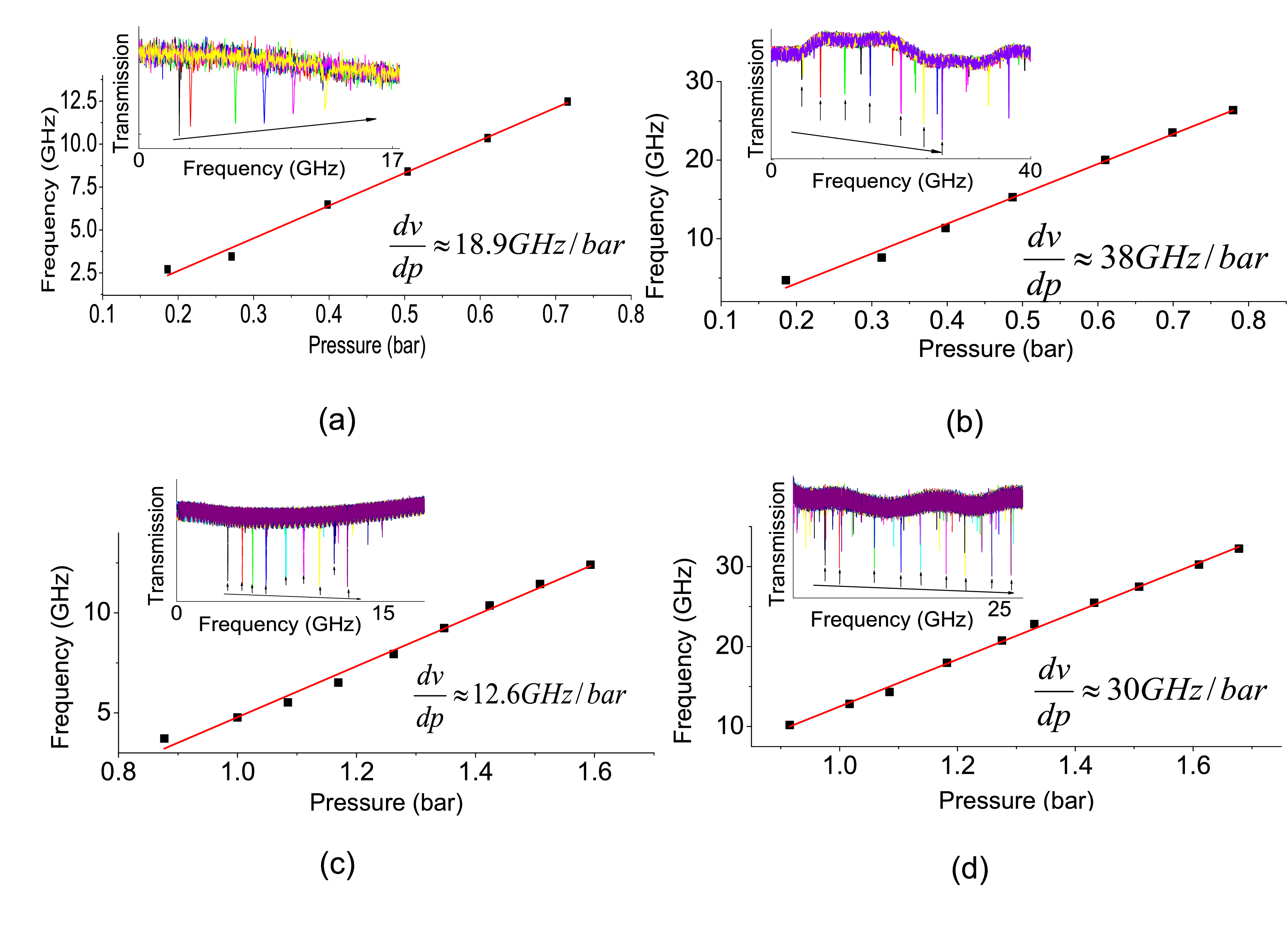}
\caption{Pressure sensing sensitivity for Sample A and Sample B at two different wavelengths. Sample A at (a) 1.55 $\mu$m and (b) at 780 nm. Sample B at (c) 1.55 $\mu$m and (d) 780 nm. The insets show the transmission spectra through the fiber coupler for increasing internal pressure. The arrows are the directions of the mode frequency shifts.}
\label{sense}
\end{figure}

In the pressure sensing experiments, we sealed the dual-input microbubbles at one end and connected  the other end to a compressed air source. The pressure was manually tuned with a pressure regulator and a calibrated electronic pressure sensor was used to record the applied pressure (see Fig. \ref{setup}). Figs. \ref{sense}(a) and \ref{sense}(b) show the relative frequency positions of a specific mode for different inner pressures in Sample A fitted with a linear plot. The slope of the fit yields the sensitivity of this microbubble to be 19 GHz/bar at 1.55$\mu$m and 38 GHz/bar at 780 nm. These values are close to the theoretical sensitivities determined from the material elasto-optical coefficient, the bulk and shear moduli, $R$, and $t$ \cite{Henze2011,Yang2015}, which are 21.5 GHz/bar at 1.55 $\mu$m and 42 GHz/bar at 780 nm. The resolution for pressure sensing can be calculated on dividing the sensitivity by the linewidth, yielding a minimal value of 1.3 mbar at 1.55 $\mu$m. At 780 nm, the sensitivity increases twofold and the linewidth is 10 times narrower. However, due to the modal coupling, the total linewidth should be the sum of the two peaks (shown in Fig. \ref{qmes}(b)); hence, the resolution for Sample A is limited to 0.55 mbar. By repeating the fabrication, we made Sample B with similar geometrical parameters and there is no obvious modal coupling at 780 nm (see Fig. \ref{qmes}(d)). The sensitivity for Sample B is 30 GHz/bar, as depicted in Fig. \ref{sense}(d), and the resolution is improved to 0.17 mbar; this represents a significant resolution improvement compared to the previous work \cite{Henze2011}.

\section{Conclusion}
In summary, we have fabricated a microbubble with submicron wall thickness and a very high Q-factor. Using this device, an ultrahigh-sensitivity aerostatic sensor was demonstrated with a resolution of 0.17 mbar at 780 nm. It could be used to detect single particles in air flow, similar to that reported by Zhu \textit{et al.} \cite{Zhu2009} or quasi-droplet index sensing \cite{Yang2014} when filled with liquid in the future. The high Q-factor is also desirable for nonlinear optics and optomechanics applications with microbubbles.

\section*{Acknowledgment}
This work was supported by the Okinawa Institute of Science and Technology Graduate University. The authors thank B. Singh Bhardwaj for SEM imaging.

\end{document}